# All-Dielectric Silicon/Phase-Change Optical Metasurfaces with Independent and Reconfigurable Control of Resonant Modes

*Carlota Ruiz de Galarreta[†], Ivan Sinev[‡], Arseny M. Alexeev[†], Pavel Trofimov[‡], Konstantin Ladutenko[‡], Santiago Garcia-Cuevas Carrillo[†], Emanuele Gemo[†], Anna Baldycheva[†], V. Karthik Nagareddy[†], Jacopo Bertolotti[†] and C. David Wright[†*]*

[†] College of Engineering Mathematics and Physical Sciences, University of Exeter, Exeter EX4 4QF, UK

[‡] ITMO University, 197101 St. Petersburg, Russia

ABSTRACT: All-dielectric metasurfaces consisting of arrays of nanostructured high-refractive-index materials, typically Si, are re-writing what is achievable in terms of the manipulation of light. Such devices support very strong magnetic, as well as electric, resonances, and are free of ohmic losses that severely limit the performance of their plasmonic counterparts. However, the functionality of dielectric-based metasurfaces is fixed-by-design, i.e. the optical response is fixed by the size, arrangement and index of the nanoresonators. A far wider range of applications could be addressed if active/reconfigurable control were possible. We demonstrate this here, via a new hybrid metasurface concept in which active control is achieved by embedding deeply sub-wavelength inclusions of a tuneable chalcogenide phase-change material within the body of high-index Si nanocylinders. Moreover, by strategic placement of the phase-change layer, and switching of its phase-state, we show selective and active control of metasuface resonances. This yields novel functionality, which we showcase via a dual- to mono-band meta-switch operating simultaneously in the O and C telecommunication bands.

Optical metasurfaces offer a technologically important route towards the realization of lightweight and compact photonic devices with novel functionalities, providing an attractive path for the future replacement of conventional, bulky, optical components. *(1-5)* Since the concept of optical metasurfaces emerged, a number of novel photonic devices with exciting properties have been reported, including frequency selective absorbers, *(6, 7)* flat lenses, *(8)* polarizers, *(9)* beam steerers, *(10)* holograms and more. *(1-3)* In such devices, subwavelength resonant building blocks (often termed meta-atoms) provide a designer interface for engineering an effective permittivity ($\varepsilon_{eff}$) and also, more interestingly, an effective non-unity permeability ($\mu_{eff}$), the latter being a property not usually present in natural materials at optical frequencies. The most promising platform for obtaining a strong magnetic response at optical frequencies is currently based on all-dielectric nanophotonics, in which the meta-atoms are made of high refractive index materials, such as silicon, germanium or gallium phosphide. *(11-18)* Isolated nanoantennas (e.g. spheres or disks) made of such materials support a series of scattering resonances (usually termed Mie resonances) of both electric and magnetic types. The effective magnetic response in high-index all-dielectric nanoantennas is driven by displacement currents, rather than, as in the case of more conventional metal plasmonic metasurfaces, by conduction currents. As a result, all-dielectric nanoantennas are also practically free from Ohmic losses, leading to much higher efficiencies of operation when compared to plasmonic-based designs. In addition, the interaction of equally strong magnetic and electric dipole resonances that is enabled by the use of dielectric meta-atoms, brings about a huge range of opportunities for the manipulation of light, and vast degrees of freedom in terms of design. For example, by engineering the relative phase and amplitude of the resonances, which can be done by changing the shape of the high-index nanoantenna, one can achieve directional scattering of light (Kerker effect), *(13)* a generalized Brewster effect, *(18)* or implement high efficiency devices based on wavefront control, such as beam deflectors, *(19)* holograms or flat high numerical-aperture lenses. *(20)*

Generally, however, the functionality of all-dielectric metamaterials and metasurfaces is fixed-by-design, i.e. the optical response is fixed by the size, shape, spatial arrangement and constituent material properties of the high-index dielectric nanoantennas used. A far wider array of potential applications could be addressed if active control of the dielectric metasurface properties and/or reconfigurability capabilities could be achieved. The active control of all-dielectric metasufaces is, however, a very under-explored topic (though some interesting approaches have been made, for example by embedding structures into a liquid crystal matrix,

*(21, 22)* or the tuning of refractive index through ultrafast photoexcitation *(23)*). We here address this key omission by developing novel all-dielectric metasurfaces in which active control is achieved by embedding a switchable and tuneable chalcogenide phase-change layer within the body of high-index all-dielectric nanoantennas (here specifically silicon nanocylinders).

Chalcogenide phase-change materials (PCMs), such as the GeSbTe-based alloys, can be quickly (nanoseconds or less) *(24)* and repeatedly (up to $10^{15}$ cycles) *(25)* switched between amorphous and crystalline states (and even between intermediate phases) by appropriate thermal, optical, or electrical stimuli. *(26, 27)* Such atomic rearrangements are non-volatile in nature (as opposed to the alternative phase-change material $VO_2$ which is volatile *(28, 29)*) and result in a huge contrast in the complex refractive index, making chalcogenide PCMs very attractive for the creation of fast, energy-efficient (low-power consumption) reconfigurable optical devices and metasurfaces. *(30)* Indeed, a number of plasmonic-based metasurfaces incorporating PCMs have been reported *(31 – 38)*, along with the direct structuring of PCMs to yield dielectric metasufaces. *(39, 40)* However, as pointed out above, the successful combination of PCMs with lossless high-index dielectric nanoresonators, yielding the exciting prospect of active and independent control of both magnetic and electric Mie-like resonances, has so far not been experimentally achieved.

In this paper, therefore, we propose and experimentally demonstrate a novel, active and reconfigurable all-dielectric hybrid Silicon/PCM metasurface, based on arrays of silicon nanodisks combined with deeply subwavelength-sized (~$\lambda_0$/100) inclusions of the chalcogenide PCM $Ge_2Sb_2Te_5$ (GST). We show experimentally, for the first time to our knowledge, how our approach can provide independent and reconfigurable control of the characteristic resonances associated with Mie-like electric and magnetic dipole modes, opening up a new route towards the design and realization of reconfigurable metasurfaces with new functionalities. This independent and reconfigurable control is achieved by positioning the GST inclusions at the electric field anti-nodes of a particular mode, and switching between amorphous and crystalline GST states to suppress that mode.

As a proof of concept, we have designed, fabricated and characterized a prototype device working in the near-infrared and consisting of a reconfigurable dual-to mono-band meta-switch that works simultaneously in reflection and transmission. Our meta-device has dual-band filtering capabilities when the GST layer is amorphous, but switches to a mono-band configuration after crystallization due to independent suppression of the electric dipole mode. Experimental results show very good agreement cf. simulations, with device efficiencies of over 80% and modulation depths exceeding 70% for hybrid Si/PCM nanodisk arrays with an embedded GST layer of just 15 nm in thickness. Furthermore, the reflectance and transmittance spectra of our novel meta-devices show robustness against angle of incidence effects under transverse electric excitation, but can also have angular filtering capabilities due to excitation of leaky Bloch modes under transverse magnetic incidence, which could well lead to useful additional applications such as de-multiplexing. In summary, we believe that by combining the design degrees of freedom and performance benefits of all-dielectric silicon nanoresonators with the active, dynamically-tuneable and reconfigurable properties of phase-change materials, our approach offers a new and practicable route to the realization of a next-generation of compact, high-efficiency, low-power consumption and active photonic devices.

**Results and discussion**

An illustration of our proposed hybrid Si/PCM metasurface is shown in **Figure 1a**. It consists of a series of $Si/Ge_2Sb_2Te_5$ nanodisks arranged in a square lattice on top of a $SiO_2$ substrate. The phase-change material $Ge_2Sb_2Te_5$ was chosen as the active layer due to its well-known and attractive physical properties, in particular high optical contrast (between phases), non-volatility, fast switching time, large number of switching cycles, and low optical losses in the near infrared when amorphous (though we note that GST and other PCM compositions work very effectively from the visible to the long infrared, and even beyond). *(24, 26, 30-33)* Moreover, as shown in **Figure. 1b**, the complex refractive index of amorphous GST matches very well the refractive index of (amorphous) silicon in the $\lambda$= 1300 nm to $\lambda$= 1600 nm window, a technologically important spectral band commonly used in optical telecommunications. After crystallization of the GST, however, an increase of the refractive index *n* and absorption coefficient *k* takes place, with an overall increment of $\Delta n \sim 1.6$ and $\Delta k \sim 1.1$ in the spectral region of interest here. *(30)* Therefore, our hybrid nanodisks effectively behave as silicon-only cylinders when the GST is amorphous, but the resonant modes supported by the array (thus its optical response) can be modified on demand by switching the GST layer between its amorphous and crystalline states (**Figure 1c**).

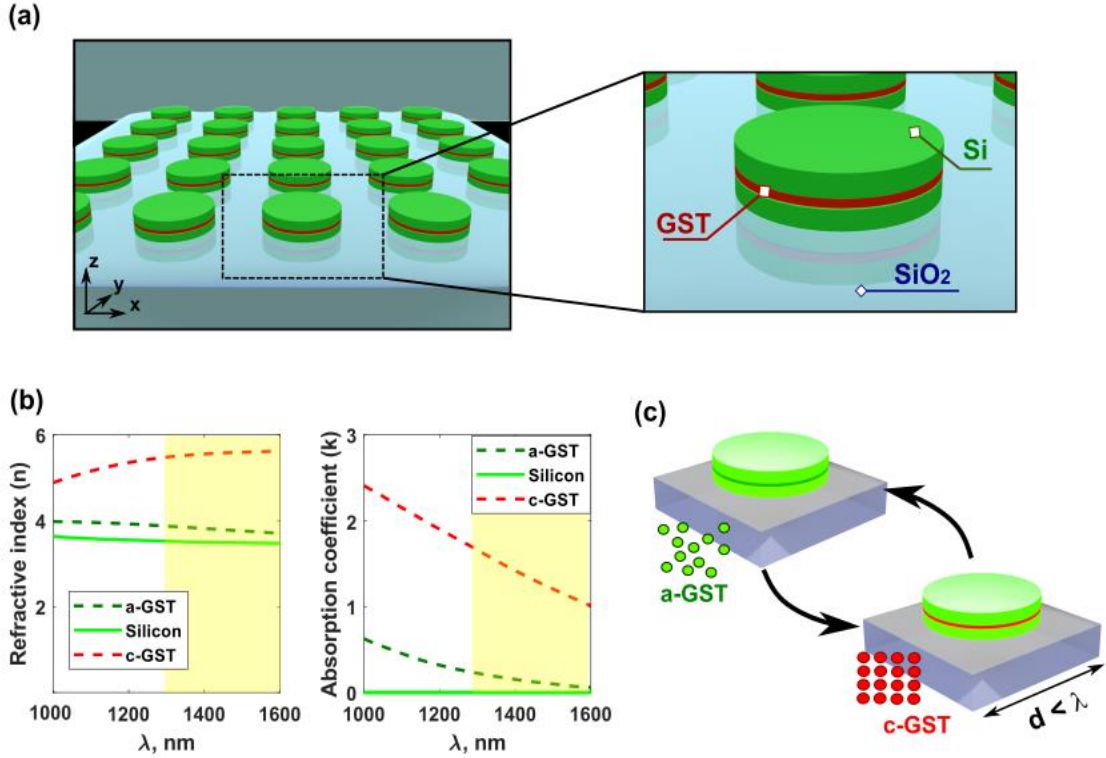

**Figure 1. (a)** Schematics of the proposed hybrid Si/PCM metasurface, consisting of arrays of Si/GST nanodisks on a SiO$_2$ substrate. **(b)** Refractive index (left) and absorption coefficient (right) of amorphous GST (a-GST), crystalline GST (c-GST) and (amorphous) silicon. The spectral region of interest is highlighted in yellow: in this region n and k of a-GST and silicon are closely matched. **(c)** Generic scheme of the device working principle: the hybrid Si/GST cylinders effectively behave as Si-only when the GST is amorphous, and the resonant modes supported by the array (thus its optical response) can be modified on demand by switching the GST layer between its amorphous and crystalline states.

The optical properties of our hybrid silicon/GST nanodisks were simulated with the finite element method (FEM), using the commercial software package Comsol Multiphysics®. **Figure 2b** shows a schematic of the generic unit cell employed in the study, which consists of a tri-layered silicon/GST/silicon nanodisk on top of a SiO$_2$ substrate. The GST thickness, $t_{gst}$, was fixed to have a maximum allowable value of 15 nm, in order to avoid the use of large PCM volumes. Keeping the PCM volume small is an especially important (and often ignored) point for successful reversible switching of the device, as re-amorphization processes require fast cooling rates (tens of degrees per nanosecond for GST) in order to quench (after melting) the PCM into its disordered amorphous phase, and such high cooling rates are difficult to achieve for thick PCM layers (due essentially to the relatively low thermal conductivity of most PCMs). *(34, 36)* Since Mie resonances are morphologically dependent, *(14 - 19)* the cylinder radius r was varied from $r$ = 140 nm to $r$ = 400 nm in steps of $\Delta r$ = 5 nm in order to gradually change the aspect ratio ($AR = t_{cyl}/2r$ with $t_{cyl}$ here fixed at 245 nm), and thus investigate its influence on the spectral position of the electric and magnetic resonances. Following the approach reported by Staude et al. *(15)*, the lattice constant $\Lambda$ was varied along with r according to $\Lambda = 2r + 220$ nm; this maintains the array in the sub-wavelength regime (therefore restricting the generation of diffraction orders) while at the same time avoids very small, and hard to fabricate, spacing between the nanodisks. Additional details on the FEM model can be found in the supplementary information section S1.

The simulated reflection and transmission coefficients of the device, under normal incidence excitation, are shown in **Figure 2b**, for the spectral range 1000 to 1700 nm, as a function of cylinder radius *r* and with the GST in the amorphous phase. The results reveal the presence of resonant modes associated with both electric dipole (ED) and magnetic dipole (MD) resonances of a single nanodisk, which can be identified as, at a particular wavelength, regions of *r* in which there is an abrupt increase in reflectance and a corresponding suppression of transmittance. An overlap of both electric and magnetic modes takes place at *r* = 275 nm, $\lambda$ = 1395 nm, which results in suppression of back scattering (reflectance), with subsequent high optical transmittance. This phenomenon is known as the first Kerker condition, *(15, 41)* and occurs here when the electric and magnetic polarizabilities of the nanodisks are in phase and comparable in strength. We also note that since silicon and a-GST are optically similar (i.e. have close matching of their refractive indices in the considered spectral range), the scattering properties of hybrid Si/a-GST nanodisks as shown in **Figure 2b** are, as would be expected, in good agreement with recently published work based on silicon-only nanodisks, *(15-20)* with the exception of small residual absorption (i.e., T + R ≠ 1) due to small dielectric losses induced by the GST layer.

**Figure 2c** shows numerical results after crystallization of the GST layer, which reveal a quite different scenario. Here, the ED Mie-like resonance is attenuated, which results in the cancellation of the first Kerker condition. The nature of such behaviour can be readily explained by looking at the nanodisk near-field distributions at the frequency of ED and MD modes. **Figure 2d** (left) shows the characteristic electric field profile of an electric resonance when the GST is amorphous, with a strong enhancement of the electric field in the center of the disk (see colorbar) surrounded by magnetic current loops (red cones). The GST layer is placed in the electric field antinodes of the mode, which results in attenuation of the resonance after crystallization, essentially due to an abrupt increase of the GST refractive index and absorption coefficient (**Figure. 2d**, right). Correspondingly, **Figure. 2e** shows the typical magnetic field profile of a magnetic dipole mode, with a strong enhancement of the magnetic field in the disk center (see colorbar), surrounded by displacement currents (white cones). Here, the GST coincides with the magnetic field antinodes, and, since the magnetic permeability of GST is unitary in both phases at optical frequencies, the displacement currents show nearly no interaction with the GST in either amorphous (**Figure 2e** (left)) or crystalline (**Figure 2e** (right)) states.

The results of **Figure 2** show that our proposed hybrid Si/PCM metasurface offers exciting possibilities for the realization of dynamic, ultrafast and individual tuning of the electric and magnetic dipolar resonances. The location and thickness of the PCM layer provide additional degrees of freedom, since separate control of ED and MD resonant modes can be achieved by placing the PCM layer in the regions where the electric field of each particular mode are strongly confined (i.e., in the antinodes of E). In addition, the morphology dependent Mie-like resonances supported by our silicon/GST nanodisks provides high flexibility towards the realization of high-

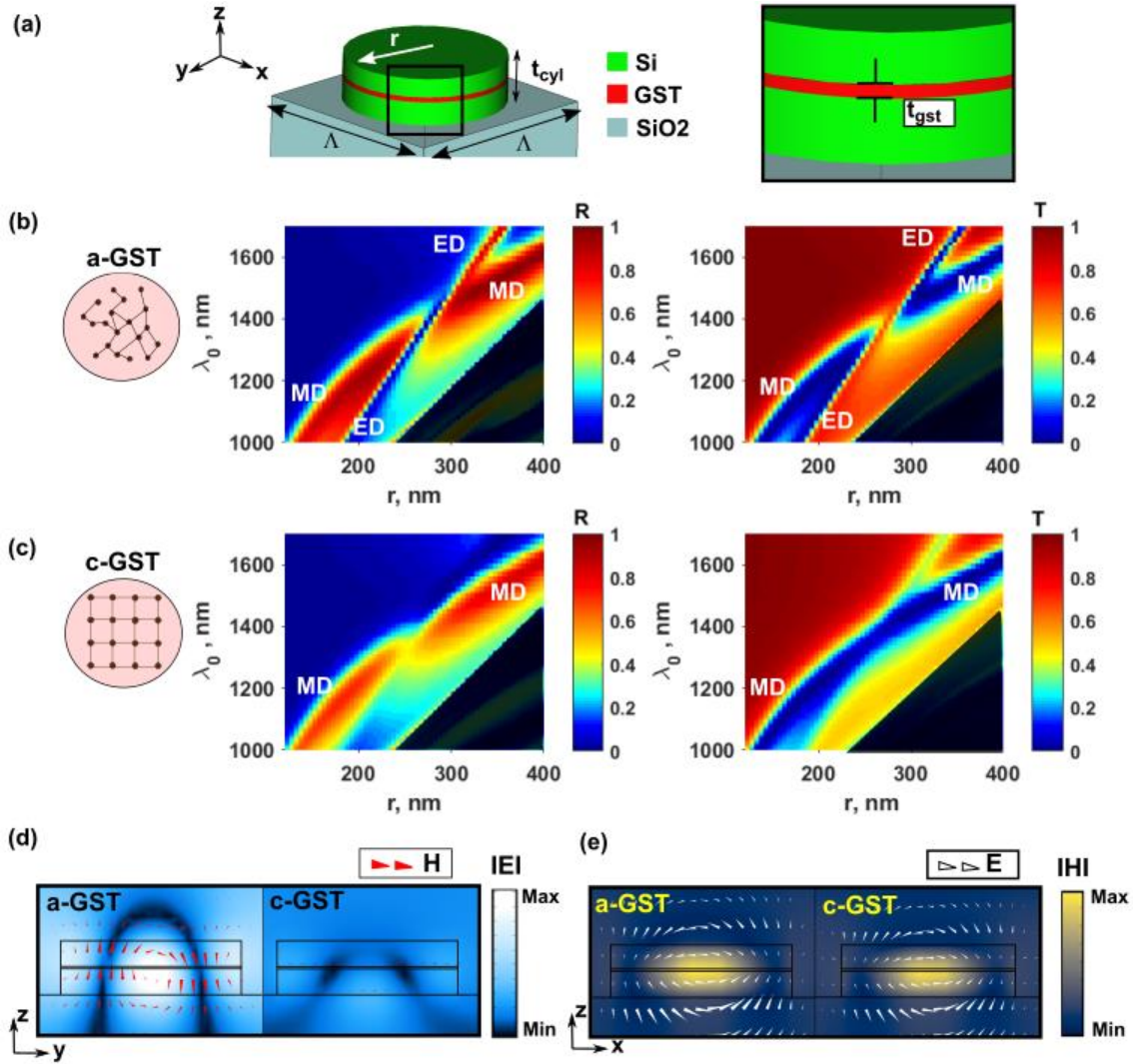

**Figure 2.** (a) Dimensions and materials of the unit cell employed in our initial FEM analyses, where $t_{cyl}$ = 245 nm, $t_{gst}$ = 15 nm, r = variable, and $\Lambda = 2r + 220$ nm. (b) Reflectance $R$ (left) and transmittance $T$ as a function of the cylinder radius $r$, showing the presence of ED and MD modes when the GST layer is amorphous. Both resonances spectrally overlap at $\lambda_0 = 1395$ nm, for a cylinder radius of $r = 265$ nm, resulting in near-zero backscattering (first Kerker condition). (c) Reflectance R (left) and transmittance T

(right) after crystallisation of the GST layer. The electric dipole is cancelled in most of the spectral region, with subsequent disruption of the Kerker condition. (d-e) Electromagnetic field distribution of the electric (d) and magnetic (e) resonances for amorphous and crystalline phases of the GST. Note that the 'black' regions in reflection/transmission spectra of (b) and (c) correspond to the diffractive regime (i.e., where light splits into different diffraction orders).

efficiency reconfigurable photonic meta-devices. For example, interference between electric and magnetic dipoles in our hybrid silicon/GST nanodisks (i.e. Kerker condition) could be exploited to selectively suppress or enhance the backscattered radiation at one particular wavelength, by changing the GST between amorphous and crystalline states. Other possible features would be the generation of a wide range of exotic tuneable spectral meta-filters, which could be realized by spectrally separating or joining both resonant modes, resulting in dual-band or broadband configurations respectively.

Having demonstrated in simulation the intriguing performance of, and prospects for, our novel, active, hybrid Si/GST optical metasurface concept, in the remainder of this paper we focus on the design fabrication and characterization of an exemplar meta-device which consists of a dual- to mono-band spectral switch/modulator with multi-channel capabilities. Here, one of the spectral bands can be selectively suppressed and modulated by switching the phase-change layer between its crystalline and amorphous states. As a proof-of-concept we design a device suitable for simultaneous and active filtering/switching in the O and C telecommunications bands (1320 nm and 1550 nm respectively). Our device consists of a square array of low aspect-ratio hybrid nanodisks, which allows one to spectrally separate the electric and magnetic Mie-like resonances, thus creating a dual-band filter. A schematic of the resulting unit cell is depicted in **Figure 3a**, where $\Lambda = 850$ nm, $t_{cyl} = 195$ nm, and $r = 666$ nm. The GST layer is placed in the middle of the nanodisk and has a thickness of $t_{gst} = 15$ nm, which was found to be thick enough for cancellation of the electric dipole after crystallization without introducing strong dielectric losses. The simulated meta-device reflectance spectra, at normal incidence and for GST layer in its amorphous and crystalline states, are depicted in **Figure 3b**. With the GST in its amorphous phase, the device shows ED and MD resonance peaks located at the desired O and C telecommunications bands, with very high device reflectances of 93% and 80% at $\lambda$ = 1320 nm and $\lambda$ = 1550 nm respectively. Crystallization of the GST layer, however, severely suppresses the ED peak due, as explained in the previous section, to the greatly increased values of n and k for the crystalline GST, which is strategically placed at the antinodes of the ED electric field distribution. Contrary to the ED case, the MD remains mostly unaffected by the GST phase transition, as also explained in the previous section.

The results of **Figure 3b** clearly show that our hybrid Si/GST metasurface can operate as an active, reconfigurable dual-band to mono-band spectral filter. With the GST layer in the amorphous state, both O and C band wavelengths (1320 and 1550 nm) are reflected, whereas switching the GST to its crystalline state results in a single-band filter in which, effectively, only the O band is reflected. The absolute modulation depth in reflectance ($MD_R = R_{am} - R_{cr}$) is 72% at $\lambda$ = 1550 nm (the modulation depth in transmittance at the same wavelength is $MD_T = T_{am} - T_{cr}$ = 65%, see supplementary information section S2A).

Finally in this section, we investigate the robustness of our device performance against changing the angle of incidence. For this purpose, we calculated reflection for a range of angles of incidence $\theta$ going from -15° to 15° in steps of $\Delta\theta$ = 0.5° for both transverse electric (TE) and transverse magnetic (TM) polarization states. **Figure 3c** and **3d** show the angular dependence of the reflectance spectra under TM excitation for a-GST and c-GST respectively. It can be seen from **Figure 3c** that the mode associated with the ED resonance of the disk is dispersionless (i.e. remains stationary when varying $\theta$), whereas the MD associated mode for non-zero angles of incidence splits into two separate resonant modes with high dispersion. The origin of the magnetic dipole splitting under TM excitation is due to different dispersion of the two-counter propagating leaky Bloch modes supported by the metasurface lattice (see the eigenmode analysis presented the supplementary information section S2B, also refs *(41, 42)*). Crystallization of the GST layer results, as expected, in the cancellation of ED mode for every angle of incidence, while the splitting of MD mode is conserved. For TE-polarized excitation (see **Figure 3e** and **3f** for a-GST and c-GST cases, respectively), both ED and MD associated modes remain unaffected by the oblique incidence while maintaining the characteristic cancellation of the ED mode after GST crystallization. These results suggest that our metasurfaces can therefore have additional features (such as tuneable multi-band filtering), upon exciting the device at different angles under TM polarization. On the other hand, angular robustness could be achieved by simply changing the incident polarization state to TE.

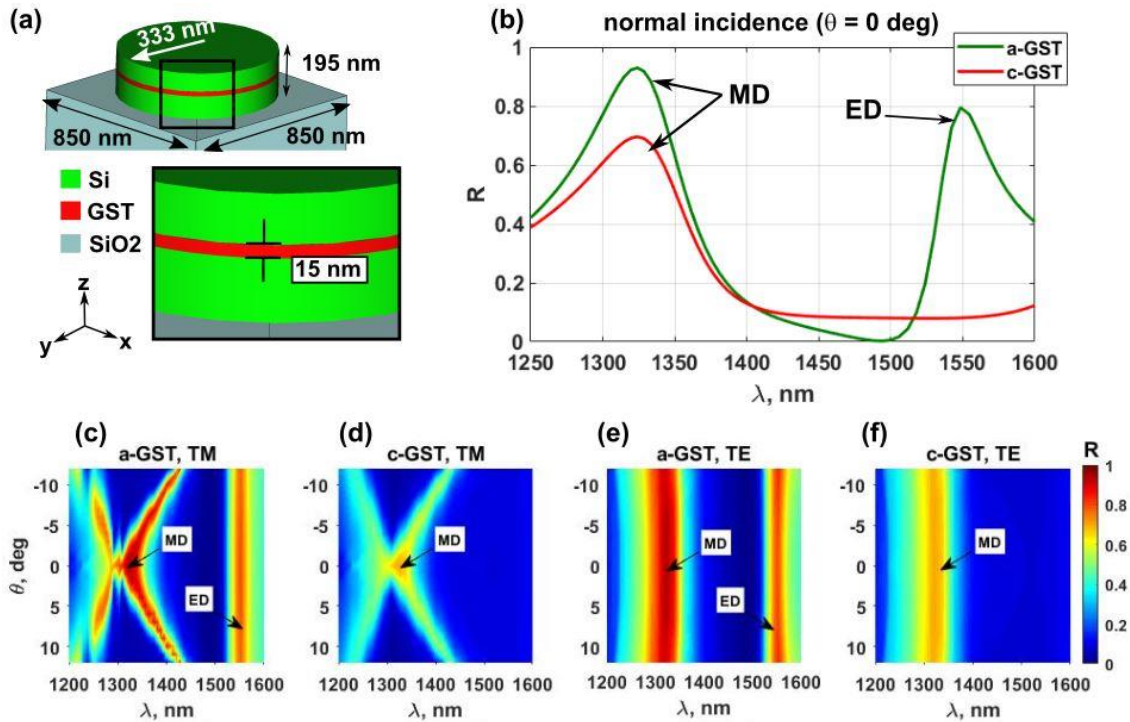

**Figure 3.** Summary of the Si/GST based meta-device. (a) Schematics and dimensions of the proposed hybrid Si/GST meta-device unit cell (b) Numerically obtained reflectance spectra at normal incidence for amorphous and crystalline states of the GST layer. (c-d) Angular reflectance under TM polarization for (c) amorphous and (d) crystalline states, showing splitting of the MD with angle of incidence, and cancellation of the ED in the crystalline phase. (e-f) Angular reflectance under TE polarization for (e) amorphous and (f) crystalline states, showing a dispersionless behaviour of the MD, and cancellation of the ED in the crystalline phase.

Our devices were fabricated in areas covering 100 μm x 100 μm using magnetron thin film sputter deposition, e-beam lithography and etching techniques, as described in the experimental section (a schematic flowchart of the whole process can be found in the supplementary material section S3). An SEM image of a typical as-fabricated device is depicted in **Figure 4a**, showing measured nanodisk diameters of 668 nm, extremely close (in fact within measurement error) of the target design diameter (as in **Figure 3a**) of 666 nm

The devices were optically characterized using back focal plane spectroscopy (method described in the experimental section and supplementary S4) to obtain the experimental reflectance spectra. Results, for normal incidence, are shown in **Figure 4b** and show excellent agreement with the simulated results previously shown in **Figure 3b**. For a-GST (green curves in **Figure 4b**), a reflectance peak of 79% corresponding to the electric dipole resonance is clearly observed at $\lambda = 1540$ nm (cf. 80% at $\lambda = 1550$ nm from numerical simulations). As expected, a second reflectance peak corresponding to the magnetic dipole is located at a shorter wavelength, experimentally at $\lambda = 1380$ nm, with a strength of R = 83% (cf. $\lambda = 1320$ and R = 93% obtained numerically). The reflectance spectrum taken after crystallization (red curve) confirms the predicted absence of ED peak at $\lambda = 1540$ nm, with the MD virtually unaffected by the phase transition. An absolute experimental contrast (modulation depth) between phases of 70% was obtained at $\lambda = 1540$ nm, very close to that predicted by our numerical models (i.e., 72% at $\lambda =$ 1550 nm). Note that here crystallization of the GST layer was achieved by thermal annealing but, as already pointed out in Section 1 above, dynamic switching (for both crystallization and amorphization) could in general be achieved by either ex-situ laser excitation (see e.g. *(26, 27, 32)*) or, more attractively, by in-situ means such as via embedded microheaters (see e.g. *(43, 44)*). Indeed, we have confirmed that the insertion of an ITO layer into our hybrid Si/PCM metadevices so as to provide a form of embedded microheater (as used successfully in *(44)* for example) while still maintaining their all-dielectric nature, does not significantly affect the optical performance (see supplementary S5).

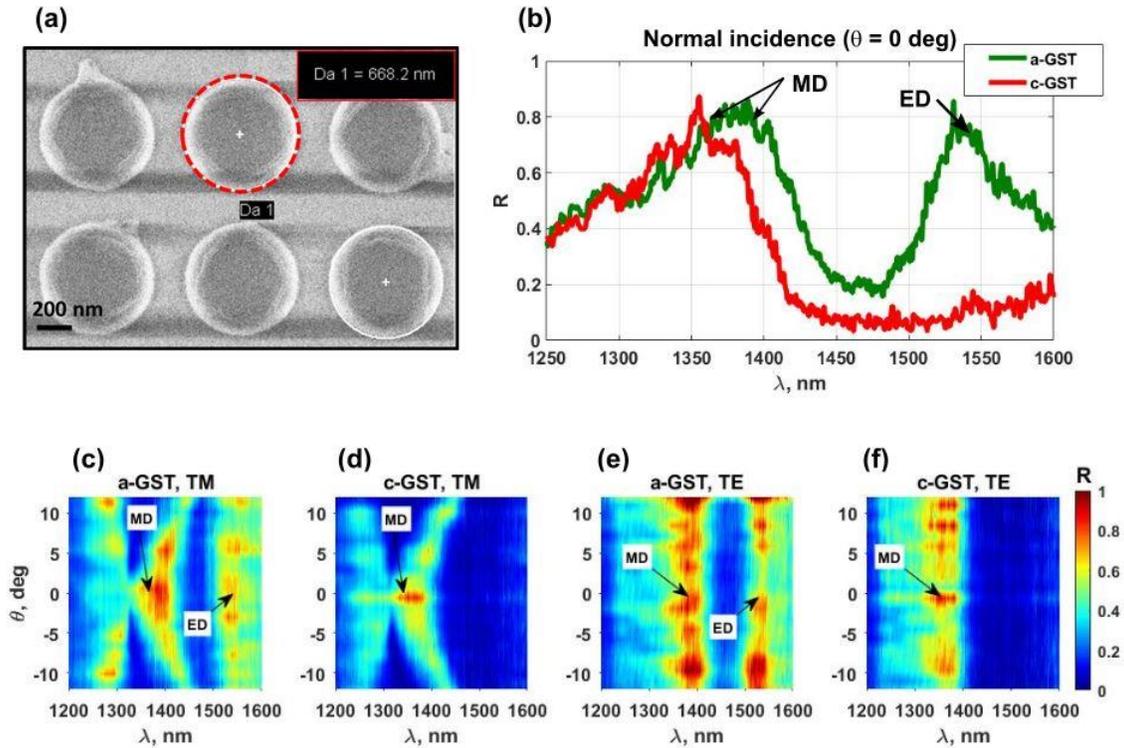

**Figure 4.** Optical response of the fabricated meta-devices. (a) SEM image of (part of) a typical as-fabricated hybrid Si/GST all-dielectric metasurface device, here showing 6 unit cells. (b) Experimentally obtained reflectance spectra for the as-fabricated device with the GST layer in both amorphous and crystalline states. (c-f) Experimentally-measured angular reflectance spectra: (c-d) Under TM excitation when the GST is (c) amorphous and (d) crystalline. (e-f) Under TE excitation for (e) amorphous and (f) crystalline GST. The experimental angular reflectance spectra show good agreement with simulation (as in Figure 3 (c-f) and confirm robustness of device performance against the angle of incidence for TE illumination, but dispersion of the MD associated mode with the angle for TM illumination.

Finally, the performance of our hybrid metasurfaces at oblique incidence was investigated under both TE and TM incidence, and compared to our numerical simulations. Thus, **Figure 4c** to **4f** show the experimentally measured reflectance spectra as a function of the angle of incidence under both TM and TE polarization and for amorphous and crystalline states of the GST layer. As predicted by simulation (see **Figure 3e** and **3f**), under TE illumination our as-fabricated metasurfaces (**Figure 4e** and **4f**) show robustness at oblique incidence (i.e. ED and MD reflection peaks remain in the same spectral position at every angle), and the ED resonance is completely suppressed for the c-GST phase over the entire range of incident angles measured (here to ±12.5 degrees). Under TM illumination, again an excellent agreement cf. simulations (**Figure 3c** and **3d**) and splitting with angle of incidence of the MD mode was observed experimentally (**Figure 4c** and **4d**), along with cancellation of the ED mode upon GST crystallization. Such angular behaviour suggests, therefore, that our hybrid metasurface devices could have additional performance possibilities, such as multichannel optical de- multiplexing (i.e. where different bands could be selectively filtered depending on the excitation angle).

**Conclusions**

We have numerically and experimentally validated a novel hybrid Si/PCM optical metasurface concept based on high-index silicon nanodisks with deeply subwavelength-sized GST inclusions (~ $\lambda_0$/100). This novel approach allows for active control and reconfigurabilty of high-efficiency, all-dielectric metasurfaces with, importantly, independent control of electric and magnetic resonances achieved by strategically locating the PCM layer only in the electric field antinodes of the mode. Our approach can be therefore exploited to create a wide range of exotic and high-efficiency photonic meta-devices with many suitable applications in the field of optical telecommunications and beyond, such as reconfigurable spectral filters, switches, absorbers or modulators, operating in reflection and/or transmission.

As a proof of concept, we proposed and experimentally demonstrated a prototype Si/PCM meta-device consisting of arrays of low aspect ratio silicon/GST nanodisks in which the magnetic and electric dipole resonances are specifically designed to be located at the telecommunication O and C bands ($\lambda$ = 1320 nm and 1550 nm respectively). This meta-device has dual-band filtering/modulation capabilities when the GST layer is amorphous, but switches to a single-band configuration after crystallization, due to cancellation of the ED mode.

Experimentally fabricated devices performed very much in line with theoretical (numerical) simulation, with an absolute maximum experimental contrast in reflection of 70% at 1540 nm obtained, using a GST layer of only 15 nm thickness. Therefore, our device has, to the best of our knowledge, the highest contrast/volume relation reported for any active PCM-based optical metasurface. The ability to use ultra-thin PCM layers in our hybrid metasurface approach is a critical factor in terms of ensuring successful re-amorphization process, which requires cooling rates of the order of tens of degrees per nanosecond, rates which are unachievable when using large PCM volumes (due to the relatively low thermal conductivity of phase-change materials). *(23, 24, 31)*

Finally, we note that while we here chose GST to implement our hybrid Si/PCM metasurface, alternative PCM compositions could of course also be used, such as the recently explored GeSbSeTe based alloys which have shown much smaller optical losses than GST (in both amorphous and crystalline states) from the near to mid infrared. *(45)* Such low-loss materials could lead to additional attractive design concepts, such as hybrid high-index/PCM all-dielectric phase-change metasurfaces for optical phase control, with potential applications including tuneable beam steerers, flat lenses and even spatial light modulators.

In summary, we believe that the active, hybrid Si/PCM optical metasurfaces proposed and demonstrated in this work can provide a practicable platform for the realization of a new generation of novel reconfigurable photonic devices, with applications in many different and important technological fields ranging from communications, to sensing, displays and much more.

**Methods:**

*Device fabrication:* Arrays of nanodisks were fabricated on 1 cm x 1 cm $SiO_2$ substrates previously cleaned with acetone and rinsed in isopropyl alcohol. First, a silicon/GST/silicon tri-layer stack was deposited using a magnetron sputtering system (Nordiko). RF sputtering in an Argon atmosphere (50 sccm) with a plasma power of 200 W was used to deposit the top and bottom silicon layers. DC sputtering was employed for the GST deposition, under the same atmosphere and a plasma power of 20 W. The chamber pressure and base vacuum for both processes were 8.5 x $10^{-2}$ Pa and 1.0 x $10^{-5}$ Pa respectively.

Next, the samples were covered with an adhesion layer (Ti-Prime) using a spinner at 4000 rpm for 20s, with subsequent post-baking at 90° C for 5 min. A negative resist (ma-N 2403) was then spin-coated at 2500 rpm for 60 s and post-baked at 90 °C for 10 min. Finally, a thin layer of conductive resist (Elecktra) was spin-coated to ease the charge dissipation during e-beam lithography (2500 rpm for 50s, post-baking at 90 °C for 2 min).

The required array pattern was then transferred to the resist via e-beam lithography (NanoBeam nB4), and subsequent development in MF-319 solution for 45 s was carried out in order to remove the unexposed areas. After lithography, the sample was post-baked at 90 °C for 5 min to increase the hardness of the remaining exposed areas.

Finally, the samples were etched in a $CHF_3$ and $O_2$ plasma mixture using an inductively coupled plasma reactive ion etching (ICP-RIE) system. ICP (300W) was used to create high-density plasma which was then accelerated towards the sample by the RIE (200W) component to achieve directional etching. A low pressure of 2 Pa was used to avoid frequent collisions inside the plasma cloud.

*Optical characterization:* Optical properties of the metasurfaces were characterized using a back focal plane spectroscopy setup (a detailed description of this setup can be found in supplementary material section S4). Sample excitation was provided by a linearly polarized beam from a supercontinuum laser source (Fianium SC400-6), which was carefully filtered and attenuated to ensure that GST inclusions do not change phase during the experiment. The beam was then focused on the sample using a 10x objective (Mitutoyo M Plan Apo NIR, NA=0.26). The end facet of a multimode fiber (50 μm core, NA=0.22) was manually scanned along the two main axes of the reflected light spot in the back focal plane of the optical setup with a step of 100 μm. Light collected by the fiber was then sent to a spectrometer (Horiba LabRAM HR800) equipped with water-cooled CCD (Andor iDus InGaAs) with sensitivity range up to 1650 nm. This provided the metasurface reflectance spectra for both TE and TM polarizations, with an angle of incidence resolution of approximately 1 degree. The spectra were normalized to the reflectance of a protected silver mirror (Thorlabs, >97.5% reflectance in the spectral range of interest).

**Supporting information:**

Section S1. FEM modelling.

Section S2. Additional FEM analyses of the hybrid Si/PCM metasurface.

Section S3. Fabrication of nanocylinder arrays.

Section S4. Experimental setup for optical characterization

Section S5. Effect of ITO layer (for use as embedded microheater) on optical performance

**Author Information:**

Corresponding Author:

\* [david.wright@exeter.ac.uk](mailto:david.wright@exeter.ac.uk)

The authors declare no competing financial interest

**Acknowledgments:**

C.D.W. acknowledges funding the EPSRC ChAMP and WAFT grants (EP/M015130/1 and EP/M015173/1). C.R.d.G. acknowledges funding via the EPSRC CDT in Metamaterials (EP/L015331/1). I.S., P.T. and A.M.A. acknowledge the support from RFBR (research project № 18-32-00527). A.M.A acknowledges support from EPSRC Impact Acceleration Account. The authors are grateful to Andrey Bogdanov for useful discussions.

# All-Dielectric Silicon/Phase-Change Optical Metasurfaces with Independent and Reconfigurable Control of Resonant Modes: Supporting information

*Carlota Ruiz de Galarreta[†], Ivan Sinev[‡], Arseny M. Alexeev[†], Pavel Trofimov[‡], Konstantin Ladutenko[‡], Santiago Garcia-Cuevas Carrillo[†], Emanuele Gemo[†], Anna Baldycheva[†], V. Karthik Nagareddy[†], Jacopo Bertolotti[†] and C. David Wright[†*]*

[†] College of Engineering Mathematics and Physical Sciences, University of Exeter, Exeter EX4 4QF, UK

[‡] ITMO University, 197101 St. Petersburg, Russia

**S1. FEM modelling**

**Figure S1a** shows a schematic of the FEM model employed in our analyses. The wavelength dependent complex refractive indices of silicon, GST and SiO$_2$ were obtained from refs. *(45)*, *(31)* and *(46)* respectively. Floquet periodic boundary conditions (PBC) were applied to the lateral faces (i.e. *x* and *y* directions) to mimic an infinite array of elements, and perfectly matched layers (PML) were placed at the top and bottom boundaries to avoid reflections from the ports. The whole structure was meshed with triangular elements in the *x* and *y* directions (having a maximum element size of $\lambda_0/10$, where $\lambda_0$ corresponds to the shortest wavelength of the analysed spectral range). Meshing in the z direction was carried out by projecting elements from the *x* and *y* directions. The SiO$_2$ substrate was assumed to be semi-infinite, thus Fresnel transmission and reflection coefficients of the unpatterned (back) side of the substrate were not taken into account.

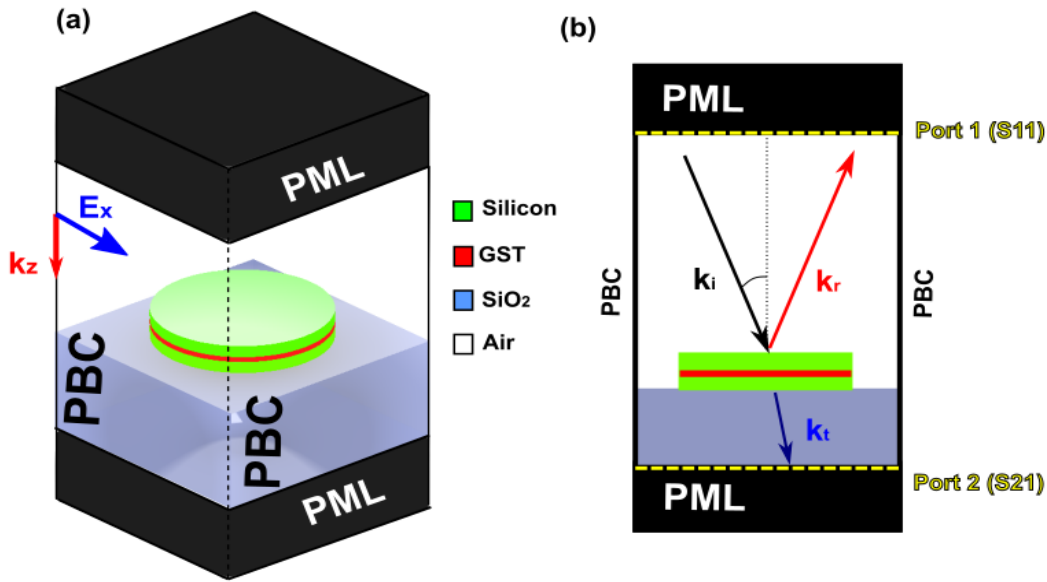

**Figure S1. Schematics of the FEM boundary conditions. (a)** Three-dimensional diagram of the FEM model, showing the system coordinates, material distribution and boundary conditions. **(b)** Two-dimensional diagram of the FEM, showing excitation and listener ports.

To obtain the reflectance and transmittance spectra, a stationary study in the frequency domain was carried out using the RF module from Comsol Multiphysics ®. The structure was excited from the air boundary at a range of angles of incidence going from -15 deg to 15 deg, as shown in **Figure S1b**. Results were displayed as a transfer matrix where the frequency dependent amplitude and phase of the reflected and transmitted waves could be extracted (i.e. modulus and argument of the S11 and S21 parameters respectively).

## S2. Additional FEM analyses of the hybrid Si/PCM metasurface

*A.     Transmittance and absorbance*

**Figure S2a** shows the transmittance spectra of the hybrid Si/GST metasurface for amorphous and crystalline states at normal incidence. It can be seen that a huge modulation depth can be achieved simultaneously in transmission (with a 65% of contrast at $\lambda_0 = 1550$ nm) and in reflection (72% in reflection as shown in the main text), which is a direct results of cancellation of the resonant electric dipole mode. Virtually zero contrast can be observed at $\lambda_0 = 1320$ nm, confirming once again that the magnetic dipole mode remains nearly unaffected by the GST crystallisation.

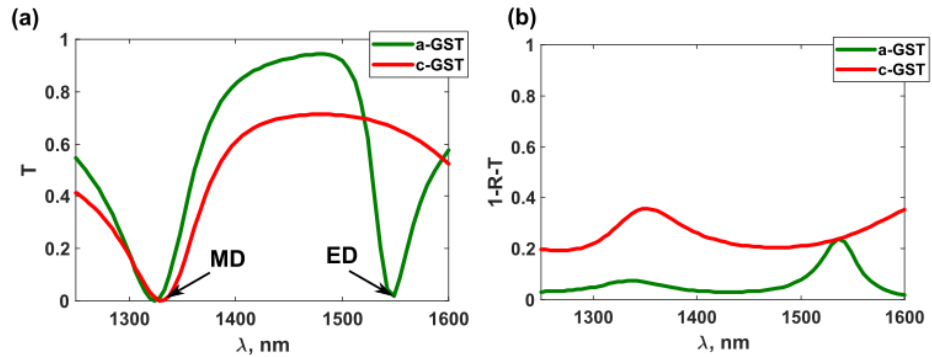

**Figure S2. Transmittance and absorbance spectra of our meta-device. (a)** Transmittance spectrum of the proposed device for amorphous and crystalline states **(b)** Absorbance spectrum for amorphous and crystalline states.

**Figure S2b** shows the absorbance spectra $A = 1 - R - T$ for the GST amorphous and crystalline states. In spite of the fact that the losses for amorphous GST are relatively low ($k =$ 0.07), it can be seen that an absorbance of 24% takes place at $\lambda_0$ = 1550 nm for the a-GST case. This is due to the interaction of the GST layer with the electric field antinodes of the electric dipole resonance (see **Figure 2d** of the main text). After crystallization, the maximum absorbance occurs at the spectral position of the magnetic dipole ($\lambda_0$ = 1320 nm) with a maximum value of 35%. Here, the electric field interactions with the GST layer are weaker (*cf.* the electric dipole case), however, the absorption coefficient $k$ is much higher for the crystalline phase ($k$= 1.7) which results in dielectric losses comparable to the amorphous phase at $\lambda_0$ = 1550 nm.

B.    *Bandstructure analysis*

Bandstructure analysis for the proposed metasurface design was performed, in Lumerical *FDTD* Solutions, to obtain additional information on the behaviour of the modes excited in the system. The results being shown in Figure **S3a** and reveal that the longer wavelength electric dipole mode is dispersionless, and is almost equally pronounced in TE and TM polarizations. The shorter wavelength magnetic dipole mode, on the other hand, exhibits different behaviour. For TE polarization, the disks remain uncoupled even for non-zero angle of incidence and the mode has low dispersion. For TM polarization, however, the disks couple strongly with each other which results into splitting of the mode into two counter-propagating Bloch modes as shown in **Figure S3b**. The higher frequency branch of the Bloch mode also exhibits anti-crossing with an electric quadrupole mode which is otherwise weakly coupled to free-space light.

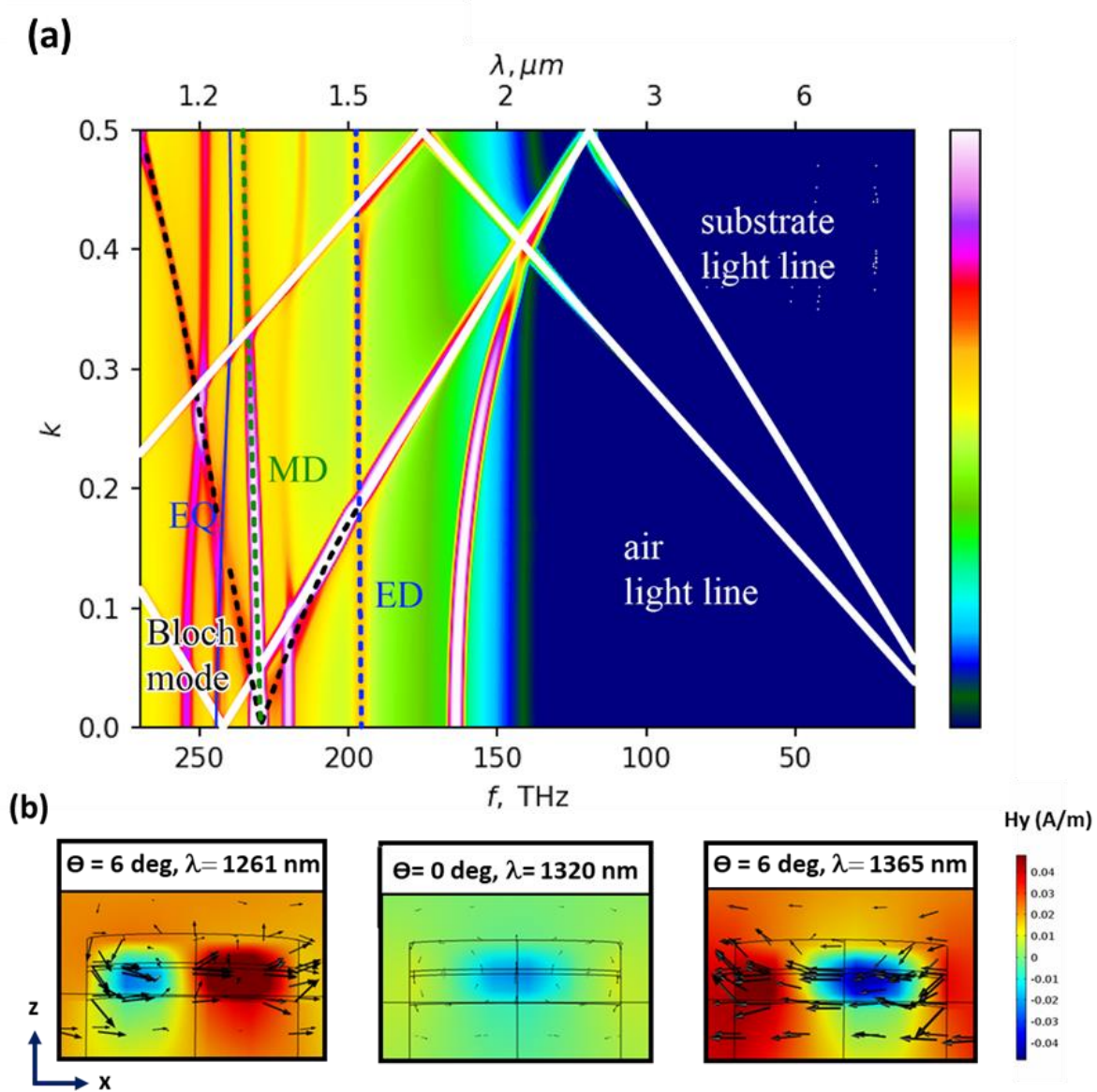

**Figure S3. Band structure and mode analysis. (a)** Bandstructure of the hybrid Si/PCM metasurface studied in the main text calculated in Lumerical FDTD. The vertical axis units are normalized to the width of the first Brillouin zone. Modes that are manifested in the measured and calculated reflection spectra are marked with dashed lines. The light lines in air and $SiO_2$ are marked with solid white lines. **(b)** Distribution of the Magnetic field y-component (colorbar) under different excitation conditions. The presence of leaky Bloch modes is confirmed by the Poynting vector (black arrows), which shows energy transport in opposite directions at oblique incidence (here 6 deg, $\lambda_0$= 1261 nm and $\lambda_0$ = 1365 nm), and a standing wave at normal incidence ($\lambda_0$ = 1320 nm).

## S3. Fabrication of nanocylinder arrays

The fabrication process for the hybrid Si/GST nanocylinder arrays is shown schematically in **Figure S4**.

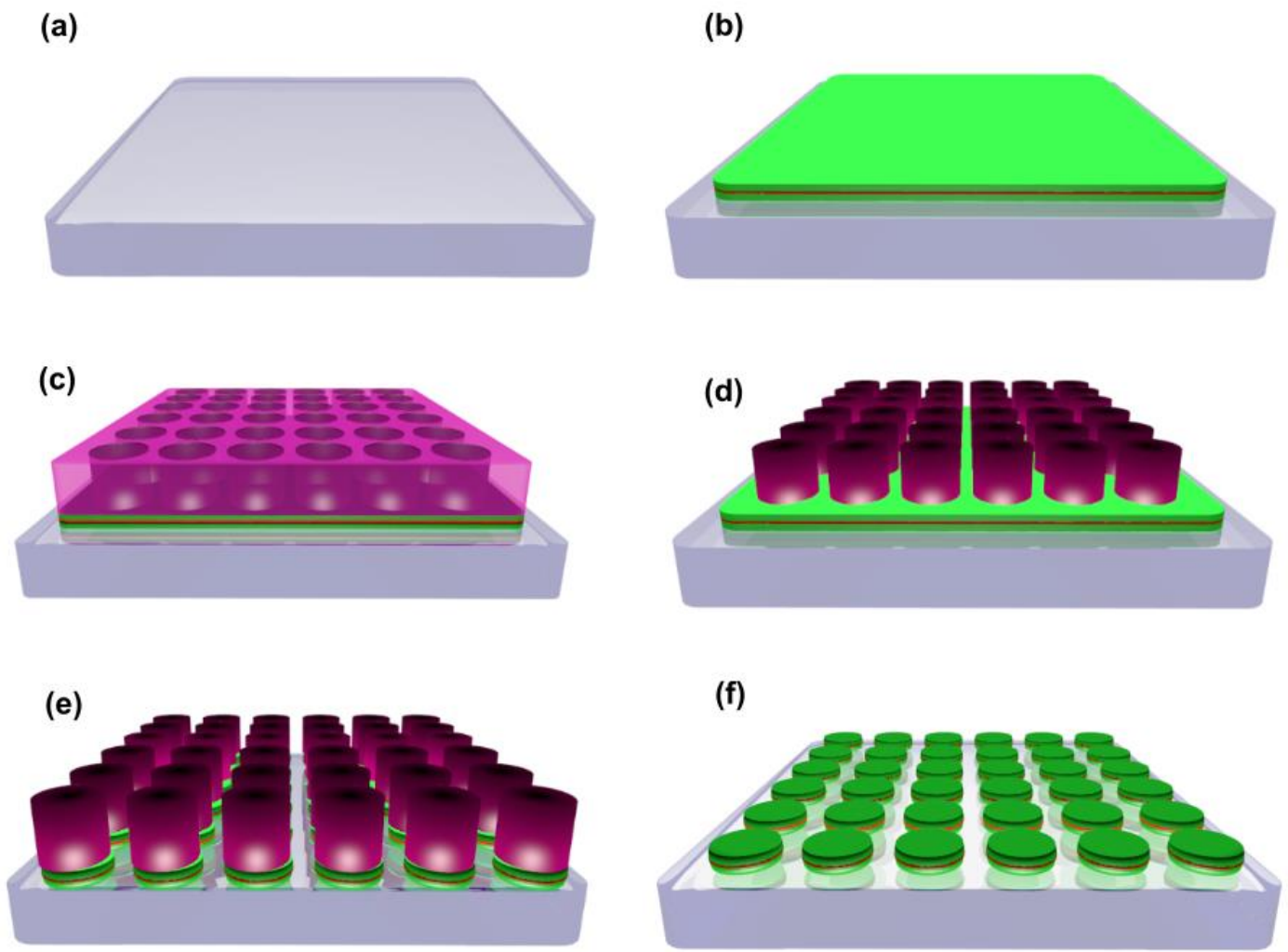

**Figure S4. Nano-fabrication process. (a)** Substrate cleaning. **(b)** Deposition of silicon, GST and silicon layers. **(c)** Spin coating of the adhesion layer, negative photoresist and conductive layer and subsequent e-beam exposure. **(d)** Development of the pattern. **(E)** Reactive ion etching. **(f)** Photoresist removal.

## S4. Experimental setup for optical characterization

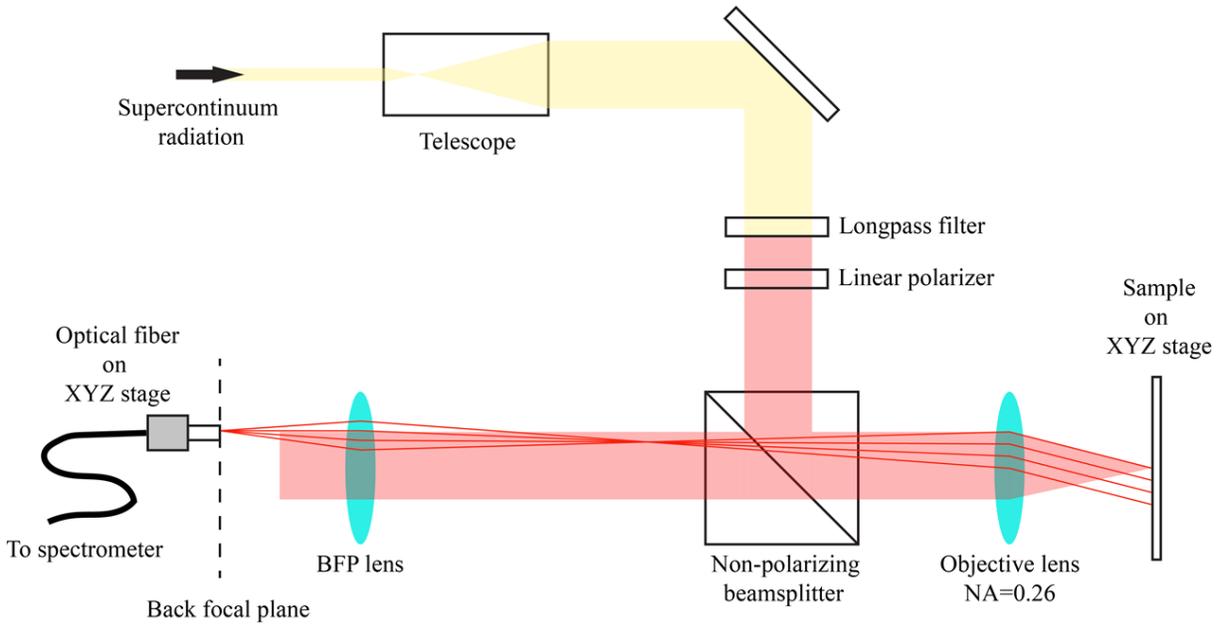

**Figure S5. Experimental setup for obtention of the angular resolved reflectance spectra.**

## S5. Effect of ITO layer (for use as embedded microheater) on optical performance

As pointed out in the main text, for dynamic switching of PCM-based metasurface devices, such as the hybrid Si/GST system developed here, excitation using ex-situ laser pulses can be used, as has been successfully demonstrated previously using both scanning and femtosecond laser approaches (see e.g. *(31, 32 47,48)*). A more practicable approach for real-world applications would however be some form of in-situ switching. Perhaps the most promising approach for such in-situ switching in all-dielectric metasurface devices is to include an ITO layer into the structure to provide a form of embedded microheater. The use of microheaters to both crystallize and amorphize GST and other phase-change materials has already been experimentally demonstrated (see e.g. *(42, 49-50)* ), and is an approach that could be readily implemented in the hybrid Si/GST metasurface devices developed in this work. For example in **Figure S6a** we show the incorporation of a thin (20 nm) ITO layer into the basic hybrid Si/GST unit cell, while in **Figure S6b** we show schematically how such a layer might be used to provide a form of in-situ microheater. In **Figure 6c**, we show (via FEM simulation) that the inclusion of such an ITO layer for the provision of an embedded microheater for in-situ switching of the GST layer, has negligible impact on the optical performance of our device. Moreover, thermal calculations confirm that both crystallization and amorphization (melting) temperatures in the GST layer would be obtainable using such ITO microheaters, and with switching energies of the order 100 pJ per cell.

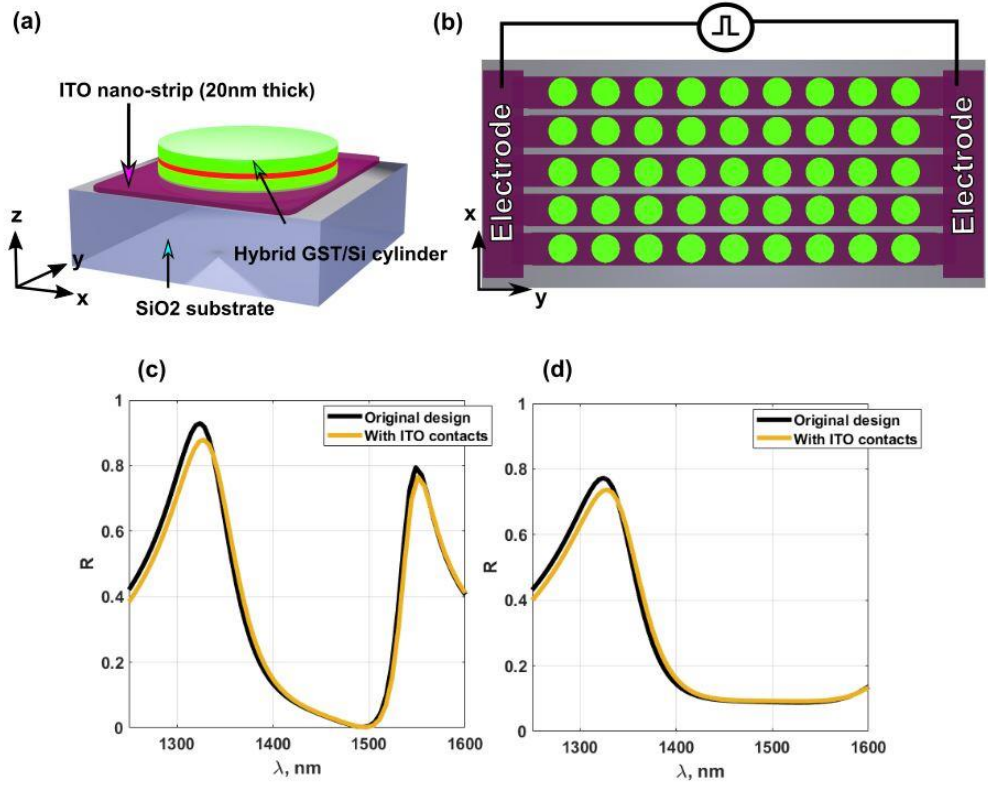

**Figure S6. In-situ electro-thermal switching of the metasurface. (a)** Inclusion of an ITO layer into the basic unit cell of the hybrid Si/GST metasuface. **(b)** One possible configuration of use of ITO layer to provide an embedded microheater for in-situ switching of the GST layer. **(c)** The optical reflectance spectrum of the hybrid Si/GST metasurface both with and without the additional ITO layer.